\pdfoutput=1

\documentclass[fleqn]{an}

\usepackage{graphicx,times}
\usepackage{natbib}
\usepackage{amssymb,amsmath}
\usepackage{epsf,psfig}
\bibpunct{(}{)}{;}{a}{}{,}

\begin{document}

\Pagespan{886}{899}
\Yearpublication{2014}
\Yearsubmission{2013}
\Month{11}
\Volume{335}
\Issue{8}
\DOI{10.1002/asna.201312109}

\title{Determining the nature of orbits in a three-dimensional galaxy model hosting a BL Lacertae object}

\author{Euaggelos E. Zotos\inst{}\fnmsep\thanks{Corresponding author:
  \email{evzotos@physics.auth.gr}\newline}}
\titlerunning{The nature of orbits in a 3D BL Lac galaxy model}
\authorrunning{Euaggelos E. Zotos}
\institute{Department of Physics, School of Science, Aristotle University of Thessaloniki,\\
GR-541 24, Thessaloniki, Greece}

\received{2013 Oct 19}
\accepted{2014 May 14}
\publonline{2014 Oct 1}


\abstract{A three-dimensional dynamical model for a galaxy hosting a BL Lacertae object is constructed. The model consists of a logarithmic potential representing an elliptical host galaxy with a bulge of radius $c_b$ and a dense massive nucleus. Using numerical experiments, we try to distinguish between regular and chaotic motion in both 2D and 3D system. In particular, we investigate how the basic parameters of our model, such as the mass of the nucleus, the internal perturbation and the flattening parameters influence the amount and the degree of chaos. Interesting correlations are presented for both 2D and 3D dynamical models. Our numerical results are explained and supported using elementary theoretical arguments and analytical calculations. Of particular interest, is the local integral of motion which have been found to exist in the vicinity of stable periodic points. The obtained numerical outcomes of the present research, are linked and also compared with several data derived from observations.}

\maketitle

\section{Introduction}
\label{intro}

BL Lacertae (BL Lac) objects are a relatively rare subclasses of Active Galactic Nuclei (AGN). The most obvious property of BL-Lac objects is that they look like a star. BL Lac objects are very strong sources of radio and infrared emission. This synchrotron emission is generally polarized. The amount of polarization and the brightness of a BL lac object is highly variable with a rapid and erratic variability. The spectrum of a BL Lac object contains very faint emission lines, or even a total lack of them. This means, that the continuum emission from the relativistic jets is strong enough to completely overwhelm the thermal emission of the host galaxy. However, their essentially featureless spectra tell us that there is very little interstellar gas around a BL Lac object.

Recently, it was shown that may be, at least sporadically, strong gamma ray sources. It is generally accepted, that BL Lacs should be interpreted within a picture where they are associated with relativistic jets pointing in the direction of the observer. The population hosting BL Lacs should, therefore, be much more abundant than the population of BL Lacs and it was suggested that it consists of FR-I radio galaxies (see \citealp[e.g.,][]{UP95} and references therein).

In the late 70s Miller and his collaborators \citep{MH77,MFH78} showed that the light from the nucleus of the prototype of the BL Lacs had a spectrum similar to that of the small M32 elliptical galaxy. Today, it is generally accepted that BL Lac objects are the central very energetic nuclei of large and luminous elliptical galaxies. Recently high-resolution images of 24 BL Lac objects between $0.3 < z < 1.3$ taken with the NOT the ESO and the VTL telescopes have revealed many aspects regarding the complicated nature of these objects (see \citealp{HTN04}).

The family of the BL Lac objects has changed drastically in the last decades with the addition of new members. The first BL Lacs were very variable and strong radio emitters. Flaring continuum emissions was even one of their selection criteria. Since then, many objects have been selected on the basis of their broad band spectral properties found through the cross-correlation of catalogues from several frequency bands. These new BL Lacs typically have little or no data from high radio frequencies.

In a recent paper (\citealp{PC08}, hereafter Paper I) the nature of orbits in a two-dimensional BL Lac dynamical model has been studied. Moreover, the authors made a successful comparison between theoretical outcomes derived from their 2D model and observational data. The present article can be considered a continuation and an expansion of Paper I, since we use a three-dimensional dynamical model in order to investigate the properties of motion in a galaxy hosting a BL Lac object. We believe, that with the aid of Observational Astronomy and the constantly increasing new data for active galaxies we are in a position today to construct better and more realistic dynamical models in an attempt to explore and shed some light in the open issue of active galaxies.

The present article is organized as follows: In Section \ref{model} we describe the properties of our gravitational galactic model. Section \ref{2dsys} is devoted to the study of the  motion in the 2D model. In this case, we try to connect the basic parameters of the system with the evolution of the amount and the degree of chaos. These results would be used as a starting point in order to explore the more complicate 3D system. In the following Section, the 3D model is investigated and the regions of phase space corresponding either to regular or chaotic orbits are determined. In Section \ref{LocInt}, we make an attempt to explain theoretically the numerically obtained results, by defying a local integral of motion. In Section \ref{obser}, a comparison with observational data is made. We conclude with Section \ref{disc}, where a brief discussion and the conclusions of this research are presented.

\section{Description of the dynamical model}
\label{model}

Our dynamical model consists of two parts: (a) a host elliptical galaxy and (b) a BL Lac object. The host galaxy is described by the logarithmic potential
\begin{equation}
V_{h}(x,y,z) = \frac{\upsilon_0^2}{2} \ln \left(x^2 + \alpha y^2+ bz^2 -\lambda x^3 + c_b^2\right).
\label{vhost}
\end{equation}
The dynamical model (\ref{vhost}) represents an elliptical galaxy with a bulge of radius $c_b$. The parameter $\upsilon_0$ is used for the consistency of the galactic units, while $\alpha$ and $b$ describe the flattening of the galaxy along the $y$ and $z$ axes, respectively. The term $- \lambda$$x^3$, $\lambda << 1$ represents an internal perturbation and therefore, deviation from axial symmetry. Potential (\ref{vhost}) has been used successfully in several previous works in order to model a triaxial elliptical galaxy \citep[e.g.,][]{CZ11}.

For the description of the BL Lac object located at the nucleus of the elliptical galaxy, we use a spherically symmetric Plummer potential
\begin{equation}
V_{n}(x,y,z) = \frac{-G M_n}{\sqrt{x^2 + y^2 + z^2 + c_n^2}},
\label{vnuc}
\end{equation}
where $G$ is the gravitational constant, while $M_n$ and $c_n$ is the is the mass and the scale length of the nucleus, respectively. This potential has been used in the past to model the central mass component of a galaxy \citep[see, e.g.][]{HN90,HPN93,ZC13}. Here we must point out, that the nucleus is not intended to represent a black hole nor any other compact object therefore, we don't include relativistic effects.

Thus, the total potential describing the motion in this active galaxy is
\begin{equation}
V(x,y,z) = V_{h}(x,y,z) + V_{n}(x,y,z).
\label{vtot}
\end{equation}

The reason for choosing potential (\ref{vhost}) for our study is threefold: (i) the logarithmic potential describes in a satisfactory way an elliptical galaxy (see \citealp{BT08}), (ii) it is a global model and therefore, can describe the motion of stars in the entire galaxy and (iii) it was also used in Paper I. Furthermore, it is also well known from observations that galaxies hosting BL Lac objects are luminous ellipticals \citep{USO00}. On the other hand, the BL Lac object itself is well described by the spherical potential (\ref{vnuc}).

The corresponding Hamiltonian is
\begin{equation}
H = \frac{1}{2}\left(p_x^2 + p_y^2 + p_z^2\right) + V(x,y,z) = E,
\label{ham}
\end{equation}
where $p_x$, $p_y$ and $p_z$ are the momenta per unit mass, conjugate to $x$, $y$ and $z$ respectively, while $E$ is the numerical value of the Hamiltonian which is conserved. In fact, $E$ is the total energy of the test particle (star).

The equations of motion for a test particle with a unit mass are
\begin{eqnarray}
&\dot{x}& = p_x ,\ \ \ \dot{y} = p_y ,\ \ \ \dot{z} = p_z \nonumber \\
&\dot{p_x}& = -\frac{\partial V}{\partial x}, \ \ \
\dot{p_y} = -\frac{\partial V}{\partial y}, \ \ \
\dot{p_z} = -\frac{\partial V}{\partial z},
\label{eqmots}
\end{eqnarray}
where, as usual, the dot indicates derivative with respect to the time. Furthermore, the equations governing the evolution of a deviation vector $\vec{w} = \left(\delta x, \delta y, \delta z, \delta p_x, \delta p_y, \delta p_z \right)$ are
\begin{eqnarray}
&\dot{(\delta x)}& = \delta p_x,\ \ \ \dot{(\delta y)} = \delta p_y,\ \ \ \dot{(\delta z)} = \delta p_z,\nonumber \\
&\dot{(\delta p_x)}& = -\frac{\partial^2 V}{\partial x^2} \delta x - \frac{\partial^2 V}{\partial x \partial y} \delta y - \frac{\partial^2 V}{\partial x \partial z} \delta z,\nonumber \\
&\dot{(\delta p_y)}& = -\frac{\partial^2 V}{\partial y \partial x} \delta x - \frac{\partial^2 V}{\partial y^2} \delta y - \frac{\partial^2 V}{\partial y \partial z} \delta z,\nonumber \\
&\dot{(\delta p_z)}& = -\frac{\partial^2 V}{\partial z \partial x} \delta x - \frac{\partial^2 V}{\partial z \partial y} \delta y - \frac{\partial^2 V}{\partial z^2} \delta z.
\label{variac}
\end{eqnarray}

\begin{figure*}
\centering
\resizebox{\hsize}{!}{\includegraphics{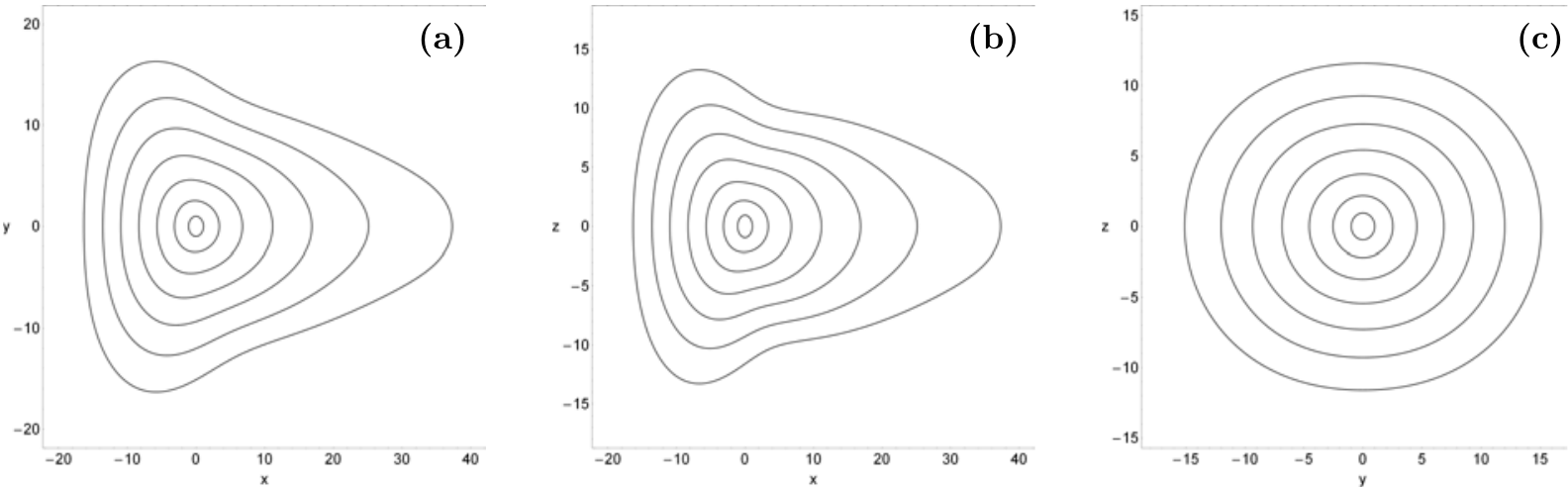}}
\caption{(a-c): Contours of the projections of the iso-density curves $\rho(x,y,z) = const$ on the $(x,y)$, $(x,z)$ and $(y,z)$ planes.}
\label{dconts}
\end{figure*}

In this work, we shall use a system of galactic units where the unit of length is 1 kpc, the unit of time is 0.9778 $\times 10^8$ yr and the unit of mass is 2.325 $\times$ $10^7$ M$_{\odot}$. The velocity unit is 10 km/s, while $G$ is equal to unity. The energy unit (per unit mass) is 100 (km/s)$^2$. In the above units, we use the following values: $\upsilon_0 = 15, c_n = 0.25$ and $c_b = 1.5$, while $\alpha$, $b$, $\lambda$ and $M_n$ are treated as parameters.

It would be of particular interest to compute the mass density $\rho(x,y,z)$ which corresponds to potential (\ref{vtot}) using the Poisson's equation
\begin{eqnarray}
\rho(x,y,z) &=& \frac{1}{4\pi G}\nabla^2 V(x,y,z) \nonumber \\
&=& \frac{1}{4\pi G} \left(\frac{\partial^2}{\partial x^2} + \frac{\partial^2}{\partial y^2} + \frac{\partial^2}{\partial z^2}\right) V(x,y,z).
\label{dens}
\end{eqnarray}
Fig. \ref{dconts}(a-c) shows the projections of the iso-density curves $\rho(x,y,z) = const$ on the $(x,y)$, $(x,z)$ and $(y,z)$ primary planes respectively, when: $M_n = 400, \alpha = 1.3, b = 1.5$ and $\lambda = 0.02$. The particular values of the contours are: (0.0017, 0.0027, 0.0045, 0.0085, 0.02, 0.07, 0.46). We can observe the clear deviation from spherical symmetry on the density distribution caused mainly by the internal perturbation $\lambda$.

For the numerical integration of the equations of motion (\ref{eqmots}) and the variational equations (\ref{variac}), a double precision Bulirsh-Stoer algorithm \citep[e.g.,][]{PTVF92} was used. The accuracy of our calculations was checked by the constancy of the energy integral (\ref{ham}), which was conserved better than one part in $10^{-11}$, although for most orbits it was better then one part in $10^{-12}$.

\section{Numerical results for the 2D system}
\label{2dsys}

Let us first study the character of orbits when the motion is restricted in the two-dimensional (2D) $(x, y)$ space (or 4D phase space), of a two degrees of freedom reduced version of the full three degrees of freedom model, where $z$ and $p_z$ are set equal to zero. Then the corresponding Hamiltonian can be written as
\begin{equation}
H_2 = \frac{1}{2}\left(p_x^2 + p_y^2\right) + V(x,y) = h_2,
\end{equation}
where $h_2$ is the numerical value of $H_2$, which is conserved. Here, $h_2$ is the total energy of the test particle moving in the $(x,y)$ plane. As the dynamical system is now two-dimensional, we can use the classical, qualitative method of plotting the successive intersections of the 2D orbits, using the $(x,p_x)$, $y=0, p_y > 0$ Poincar\'{e} Surface of Section (PSS), in order to distinguish between regular and chaotic motion. This method has been extensively applied to Hamiltonian systems with two degrees of freedom, as in these systems the PSS is a two-dimensional plot. The results obtained from the study of the 2D system will be exploited in order to help us understand and interpret the complicated phase space of the 3D Hamiltonian system, which will be presented in the following section.  	

\begin{figure*}
\centering
\resizebox{\hsize}{!}{\includegraphics{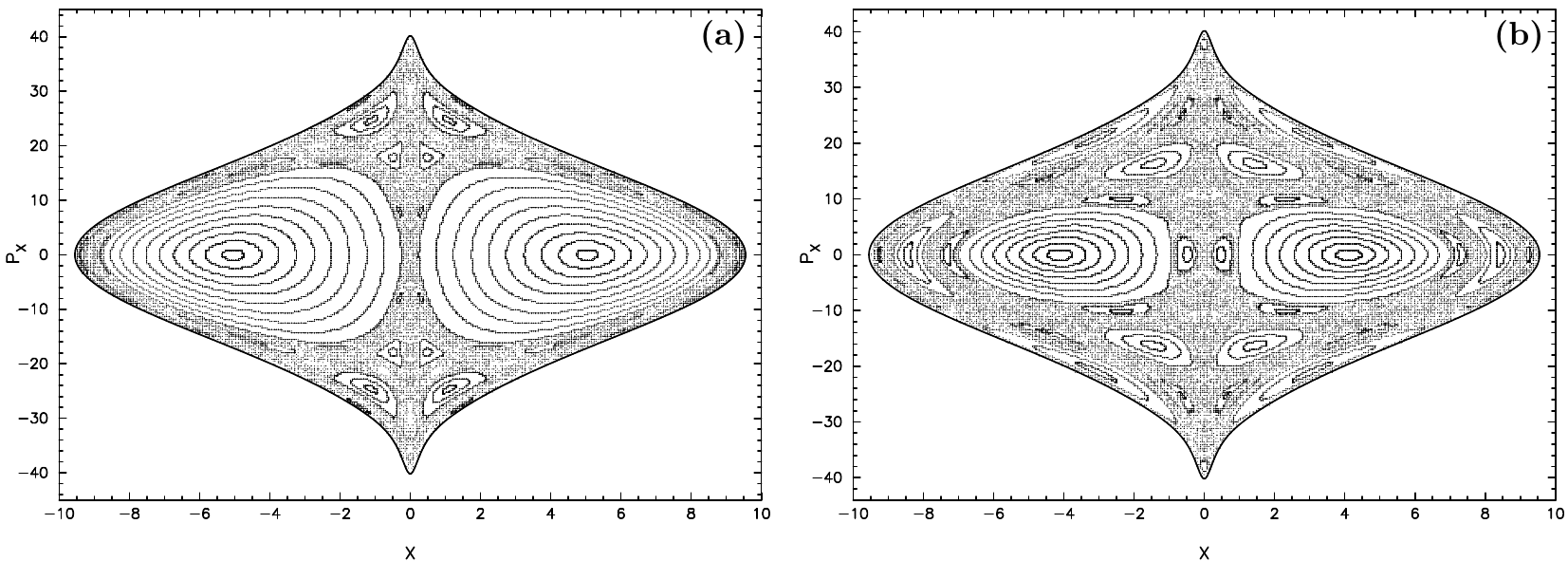}}
\caption{(a-b): The $(x,p_x)$ phase plane for the two-dimensional model when $h_2 = 500$. The values of the parameters are $\lambda = 0$, $M_n = 100$, while (a-left): $\alpha = 1.3$ and (b-right): $\alpha = 1.8$.}
\label{set1}
\end{figure*}

In Fig. \ref{set1}(a-b) we present the $(x,p_x)$ phase planes for the motion of a star in our galactic model, which was obtained by means of numerical integration of the equations of motion (\ref{eqmots}) for two different values of the flattening parameter $\alpha$, when $\lambda = 0$ and $M_n = 100$. The value of the energy is $h_2 = 500$ and remains constant so that in all phase planes $x_{max} \simeq 10$. Fig. \ref{set1}a shows the phase plane when $\alpha = 1.3$. Since $\lambda = 0$ the plot is symmetrical not only to $x$ axis but also to $p_x$ axis. Here, the majority of the phase plane is covered by initial conditions which correspond to regular orbits. In fact, there are two considerable regular regions which contain invariant curves produced by quasi-periodic orbits which are characteristic of the 1:1 resonance. Apart from these orbits, we can observe several sets of islands of invariant curves embedded in the chaotic sea which produced by other types of resonances. In particular, there are regular regions produced by quasi-periodic orbits characteristic of the 1:2, 2:3 and 3:4 resonances. Moreover, with a closer look at the phase plane, especially near the center or near the outer parts, we can distinguish tiny sets of islands of invariant curves produced by secondary resonances. The outermost solid line is the Zero Velocity Curve (ZVC) at the $(x,p_x)$ phase plane which contains all the invariant curves and it is defined as
\begin{equation}
f_1(x,p_x) = \frac{1}{2}p_x^2 + V(x) = h_2.
\label{ZVC1}
\end{equation}
In Fig. \ref{set1}b we present the case when $\alpha = 1.8$. It is evident, that the chaotic area has been increased. Furthermore, the resonant phenomena look more prominent now. Therefore, our numerical results suggest that the flattening parameter $\alpha$ affects not only the amount of chaotic orbits but also the portion of resonant orbits.

\begin{figure*}
\centering
\resizebox{\hsize}{!}{\includegraphics{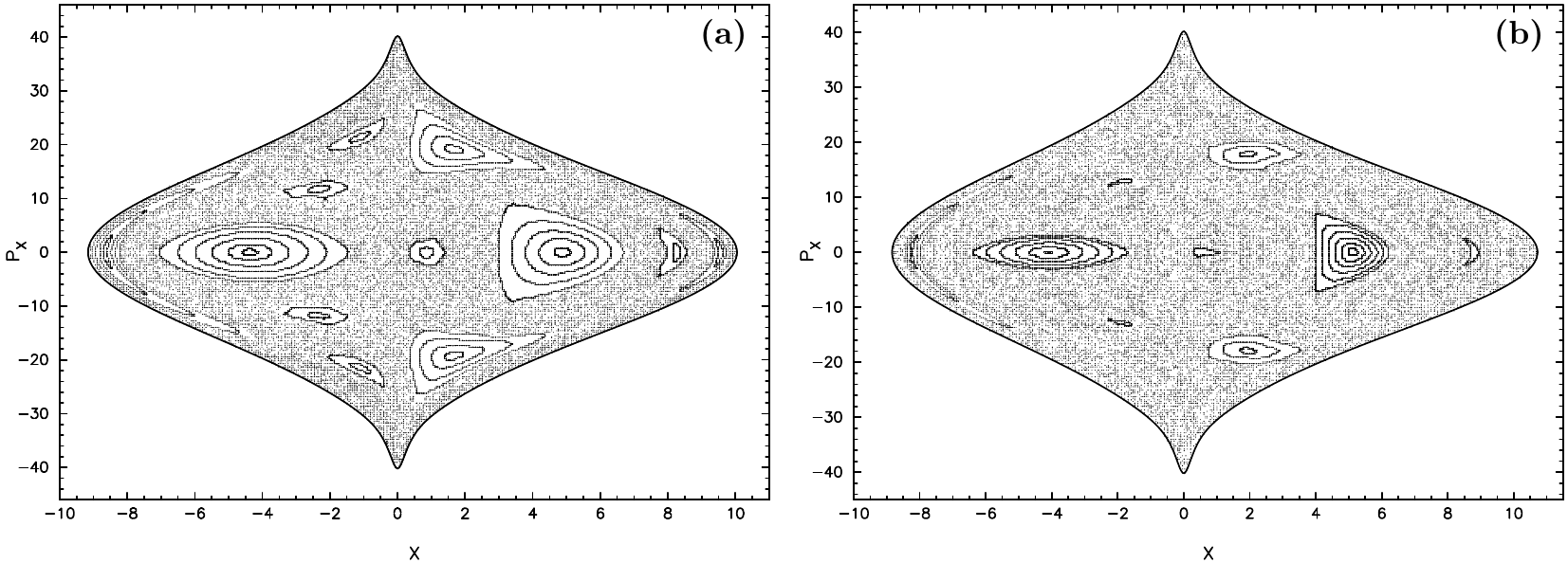}}
\caption{(a-b): The $(x,p_x)$ phase plane for the two-dimensional model when $h_2 = 500$. The values of the parameters are $M_n = 100$, $\alpha = 1.5$, while (a-left): $\lambda = 0.01$, and (b-right): $\lambda = 0.02$.}
\label{set2}
\end{figure*}
\begin{figure*}
\centering
\resizebox{\hsize}{!}{\includegraphics{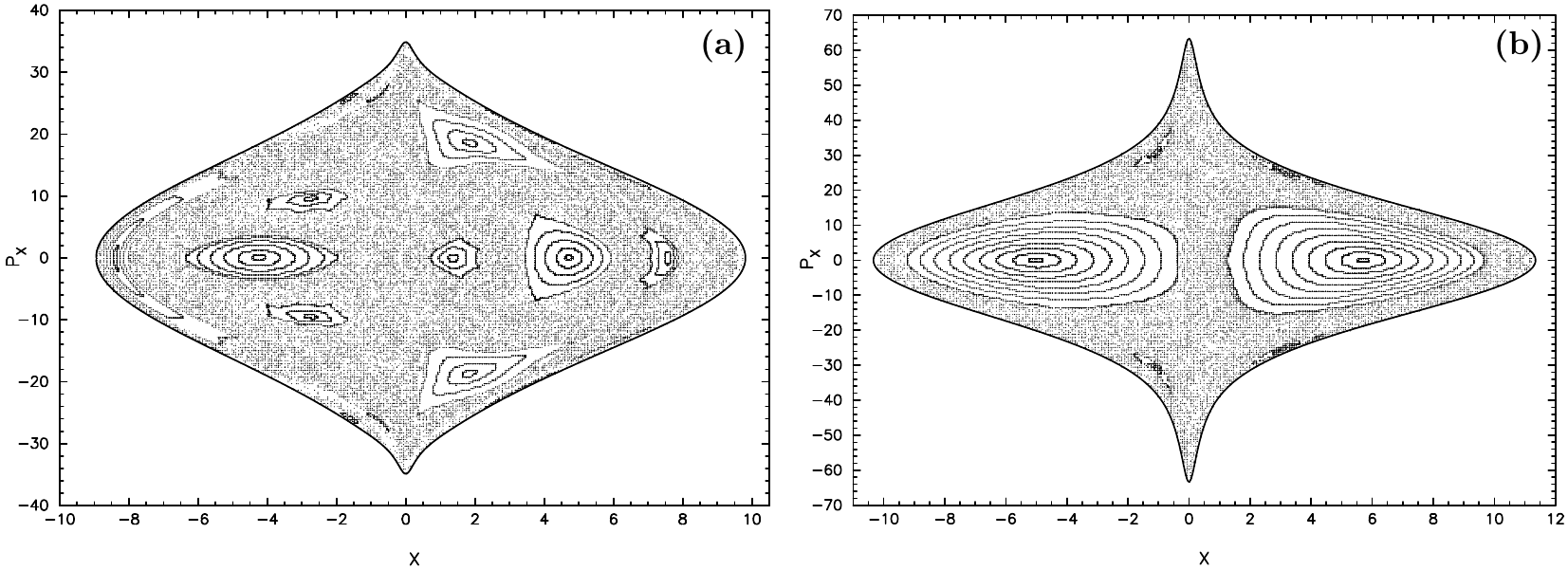}}
\caption{(a-b): The $(x,p_x)$ phase plane for the two-dimensional model when $h_2 = 500$. The values of the parameters are $\lambda = 0.01$, $\alpha = 1.5$, while (a-left): $M_n = 50$, and (b-right): $M_n = 400$.}
\label{set3}
\end{figure*}

Fig. \ref{set2}(a-b) presents the $(x,p_x)$ phase planes for two different values of the internal perturbation $\lambda$, when $\alpha = 1.5$ and $M_n = 100$. The value of the energy is again $h_2 = 500$. In Fig. \ref{set2}a where $\lambda = 0.01$ we observe that the a large unified chaotic sea exists in the phase plane. However, there are also several regions of regular motion embedded in the chaotic sea. On the other hand, in Fig. \ref{set2}b where $\lambda = 0.02$ the chaotic domain has been increased substantially and consequently regular motion is confined to small islands of invariant curves. Thus, we may conclude that the internal perturbation $\lambda$ plays a very important role on the orbital structure of the dynamical system. In fact, the stronger is the internal perturbation the more dominant is the chaotic motion.

We proceed our investigation, by presenting in Fig. \ref{set3}(a-b) the $(x,p_x)$ phase planes for two different values of the mass of the nucleus $M_n$, when $\lambda = 0.01$ and $\alpha = 1.5$. The value of the energy is once more $h_2 = 500$. When the nucleus has a relative small mass $M_n = 50$ one can identify, in the phase plane presented in Fig. \ref{set3}a, several sets of islands of invariant curves produced by resonant orbits which are all embedded in the vast chaotic sea. We observe, that Fig. \ref{set3}a is very similar to Fig. \ref{set2}a. Things are quite different in Fig. \ref{set3}b where $M_n = 400$. Here, there are mainly two large regions of regular motion inside the unified chaotic sea. A more careful inspection shows that resonant orbits are still present. The main difference between the pattern of the two phase planes is that the area of the phase plane shown in Fig. \ref{set3}b has been considerably increased because the more massive nucleus increases the velocities of the stars, particularly near the center of the galaxy. Thus, our numerical outcomes indicate that the presence of a massive and dense nucleus affects not only the percentage of the chaotic orbits in the phase plane but also the velocities of stars near the central region.

\begin{figure*}
\centering
\resizebox{\hsize}{!}{\includegraphics{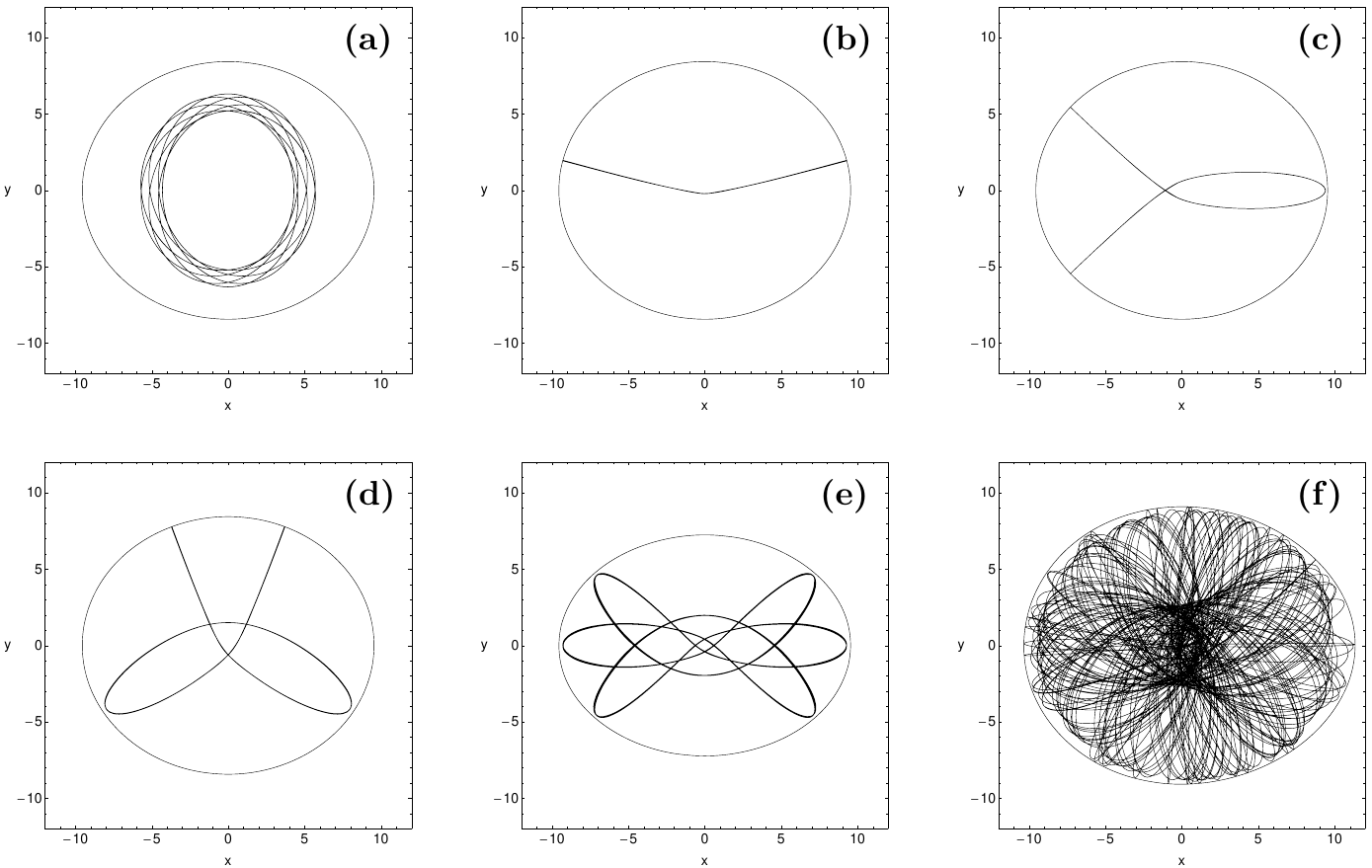}}
\caption{(a-f): Six representative orbits of the 2D dynamical system. The values of all the parameters and the initial conditions of the orbits are given in the text.}
\label{orbs2D}
\end{figure*}

Fig. \ref{orbs2D}(a-f) shows six typical two-dimensional orbits in the potential (\ref{vtot}). If Fig. \ref{orbs2D}a we see a loop orbit circulating around the center of the galaxy. The initial conditions of this orbit are $x_0 = 4.3$ and $p_{x0} = 0$. Fig. \ref{orbs2D}b shows an orbit with initial conditions $x_0 = 1.44$ and $p_{x0} = 27.85$, which is a characteristic example of the 1:2 resonance. Moreover, in Fig. \ref{orbs2D}c we present a 2:3 resonant orbit which produces a set of three islands of invariant curves in the phase plane shown in Fig. \ref{set1}a. The initial conditions are $x_0 = 9.4$ and $p_{x0} = 0$. A typical 3:4 resonant orbit with initial conditions $x_0 = 0.49$ and $p_{x0} = 17.86$ is given in Fig. \ref{orbs2D}d. Furthermore, Fig. \ref{orbs2D}e depicts a complicated quasi-periodic 3:5 resonant orbit which produces a chain of five small islands of invariant curves in the phase plane of Fig. \ref{set1}b. This orbit has initial conditions $x_0 = 9.28$ and $p_{x0} = 0$. Finally, in Fig. \ref{orbs2D}f we present a chaotic orbit with initial conditions $x_0 = 0.5$ and $p_{x0} = 0$ which correspond to the chaotic sea of the phase plane shown in Fig. \ref{set3}b. In all cases, the values of all the other parameters are as in Fig. \ref{set1}a apart form cases (e) and (f). All orbits were calculated for a time period of 100 time units, where we take $y_0 = 0$ and the value of $p_{y0}$ was found always from the energy integral (\ref{ham}). It is interesting to note that, all regular orbits do not approach the central nucleus, while the chaotic orbit passes arbitrary through the nucleus. The outermost solid curve which surrounds all orbits is the limiting curve in the $(x,y)$ plane which is calculated as
\begin{equation}
f_2(x,y) = V(x,y) = h_2.
\label{ZVC2}
\end{equation}

\begin{figure}
\includegraphics[width=\hsize]{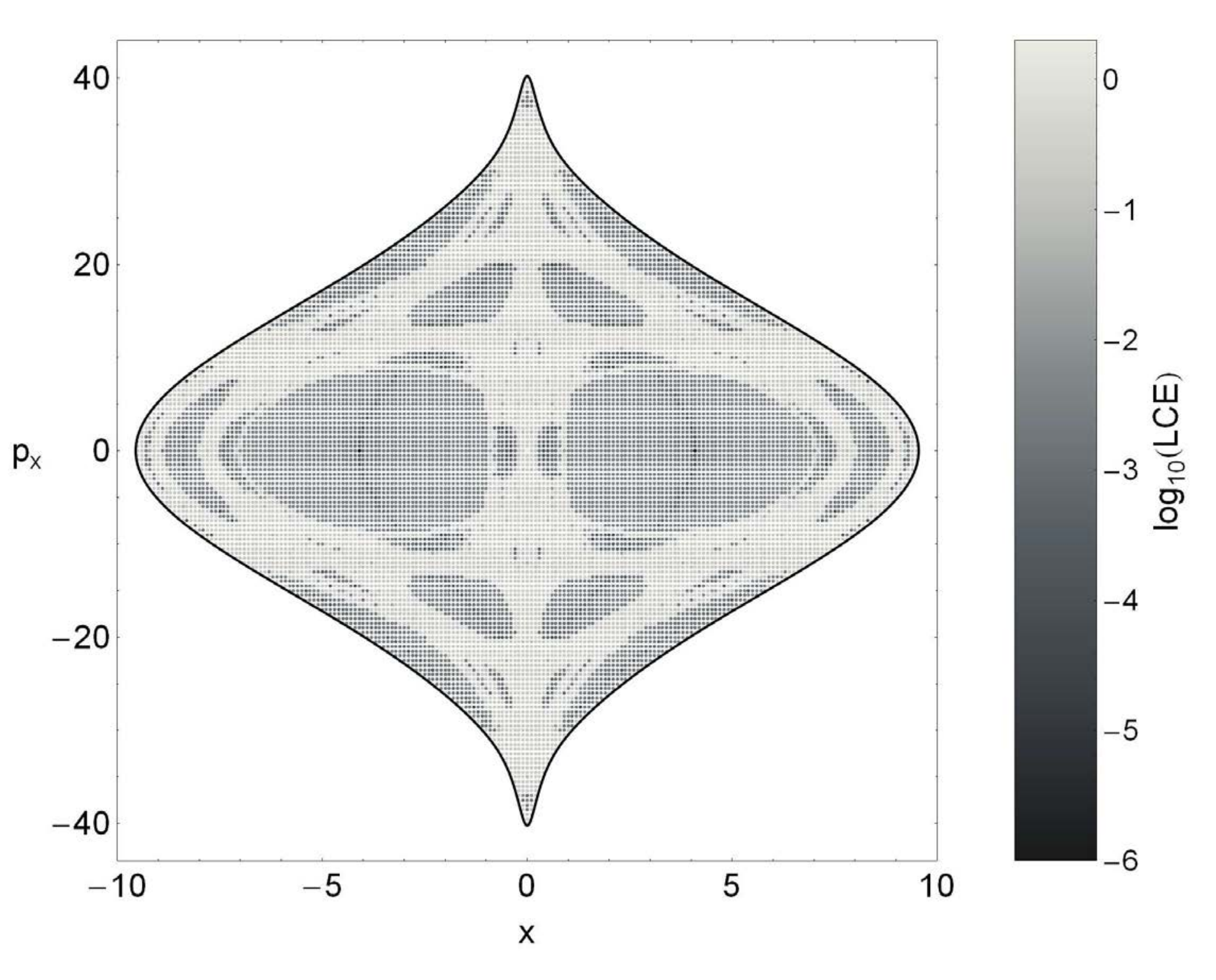}
\caption{Regions of different values of the logarithm of LCE on the PSS when $\lambda = 0$, $M_n = 100$, $\alpha = 1.8$ and $h_2 = 500$. Light grey colors correspond to chaotic motion, while dark grey colors indicate ordered motion.}
\label{grid2D}
\end{figure}

\begin{figure*}
\centering
\resizebox{\hsize}{!}{\includegraphics{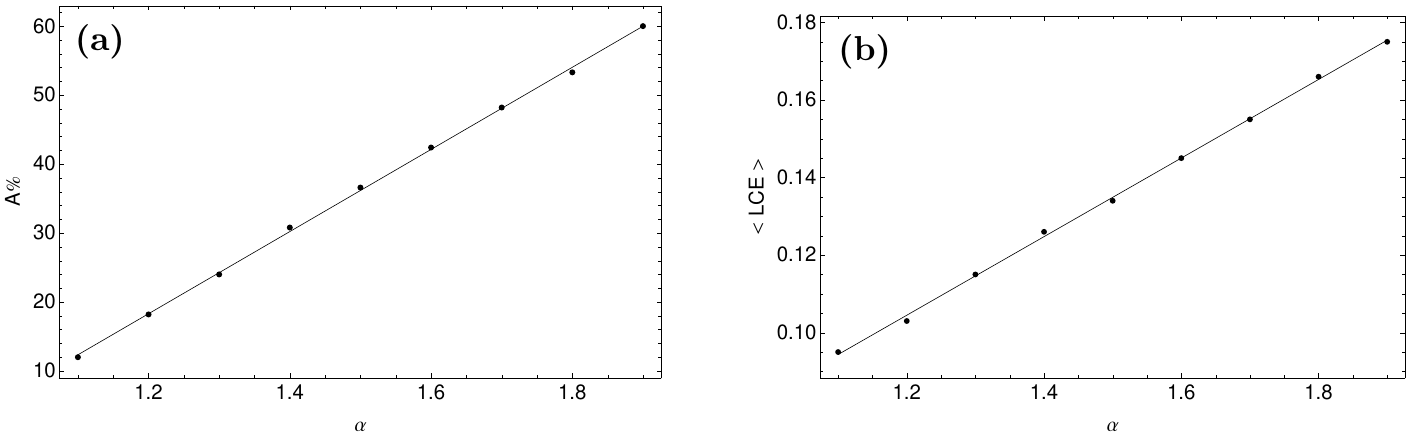}}
\caption{(a-b): (a-left): A plot of the area A\% on the $(x,p_x)$ phase plane covered by chaotic orbits as a function of the flattening parameter $\alpha$. (b-right): A plot showing the relation between the average value of LCE and $\alpha$.}
\label{aevol2D}
\end{figure*}

It would be of particular interest to connect the amount and the degree of chaos with the variable parameters of the dynamical system, that is the internal perturbation $\lambda$, the flattening parameter $\alpha$ and the mass of the nucleus $M_n$. In Fig. \ref{aevol2D}a we present the evolution of the percentage A\% on the $(x,p_x)$ phase plane covered by chaotic orbits as a function of the flattening parameter $\alpha$, when $\lambda = 0$, $M_n = 100$ and $h_2 = 500$. We observe, that the chaotic percentage increases almost linearly with increasing $\alpha$. Here we must point out, that the chaotic percentage A\% is calculated as follows: in every phase plane we construct a grid containing $10^4$ initial conditions $(x_0,p_{x0})$. Then, we integrate these orbits for a time period of 2 $\times$ $10^4$ time units distinguishing between regular and chaotic orbits by calculating the value of the Lyapunov Characteristic Exponent - LCE (see \citealp{LL92}). Thus, A\% can be obtained by dividing the number of chaotic orbits to the total number of tested orbits. Such a dense grid of initial conditions on the PSS when $\alpha = 1.8$ is presented in Fig. \ref{grid2D}. The values of the logarithm of the LCE are plotted by different shades of grey. In Fig. \ref{grid2D} we clearly distinguish between light grey regions, where the motion is chaotic and dark grey regions, where it is ordered. In order to have an estimation regarding the degree of chaos in our 2D dynamical system, we have also computed the average value of the LCE of the chaotic orbits in each phase plane. Fig. \ref{aevol2D}b shows a plot of the evolution of $< \rm LCE >$ as a function of $\alpha$. Again, we observe a linear trend. Combining the results presenting in Fig. \ref{aevol2D}(a-b) we may say, that the relation between the flattening parameter $\alpha$ and both the amount and the degree of chaos is linear.

\begin{figure*}
\centering
\resizebox{\hsize}{!}{\includegraphics{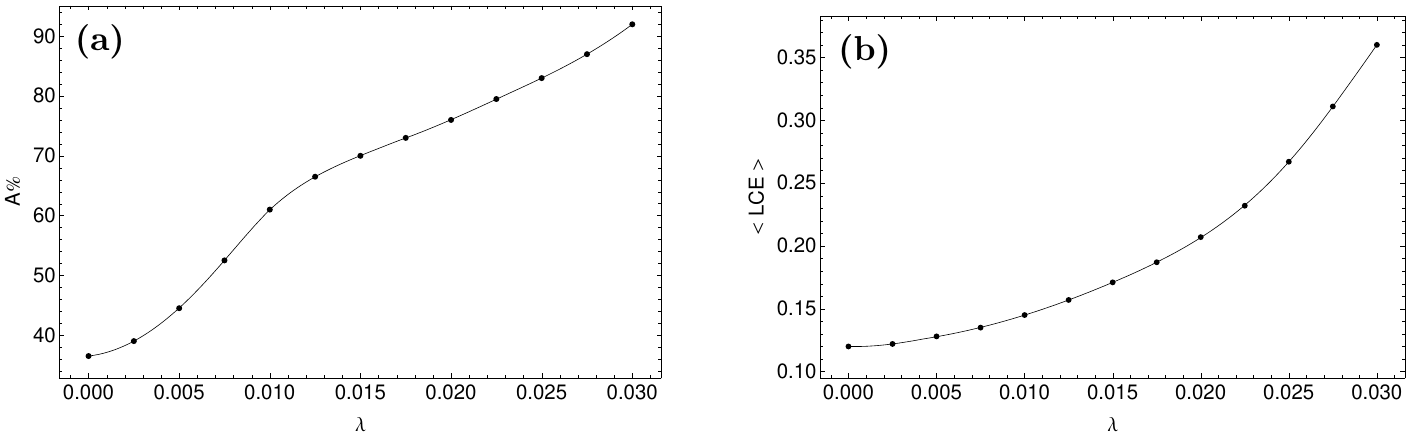}}
\caption{(a-b): (a-left): A plot of the area A\% on the $(x,p_x)$ phase plane covered by chaotic orbits as a function of the internal perturbation $\lambda$. (b-right): A plot revealing the relation between the average value of LCE and $\lambda$.}
\label{lamdaevol2D}
\end{figure*}

\begin{figure*}
\centering
\resizebox{\hsize}{!}{\includegraphics{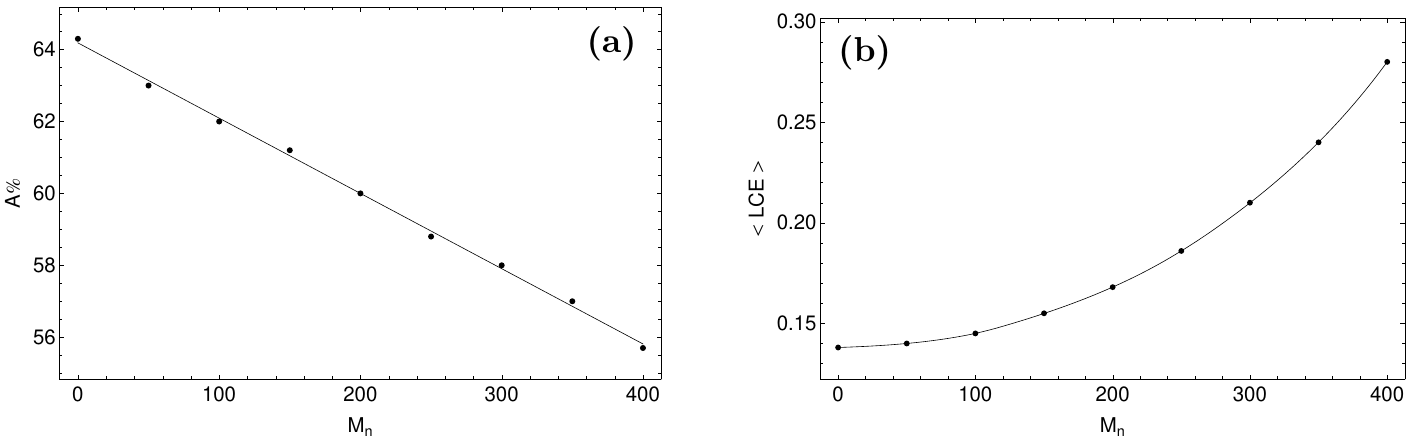}}
\caption{(a-b): (a-left): A plot of the area A\% on the $(x,p_x)$ phase plane covered by chaotic orbits as a function of the mass of the nucleus $M_n$. (b-right): A plot depicting the relation between the average value of LCE and $M_n$.}
\label{Mnevol2D}
\end{figure*}

Following the same philosophy as above, we investigated the evolution of the chaotic percentage A\% on the $(x,p_x)$ phase plane as a function of the internal perturbation $\lambda$, when $M_n = 100$, $\alpha = 1.5$ and $h_2 = 500$. Our results are given in In Fig. \ref{lamdaevol2D}a. We see, that for small values of $\lambda$ $(\lambda < 0.01)$ the chaotic percentage increases rapidly, while for larger values of $\lambda$ it follows an almost linear increase. On the other hand, the plot in Fig. \ref{lamdaevol2D}b indicates that the evolution of $< \rm LCE >$ with respect to $\lambda$ follows a monotone rapid increase. Thus, it is evident that the larger the value of the internal perturbation the stronger is the observed chaos in the dynamical system. Last but not least, Fig. \ref{Mnevol2D} (a-b) shows the influence of the mass of the nucleus to the chaotic orbits. Looking at Fig. \ref{Mnevol2D}a one may assume that as the value of the mass of the nucleus increases the chaotic percentage decreases following a linear trend. However, this is not entirely correct. In fact, what we see in Fig. \ref{Mnevol2D}a is only a numerical artifact which does not correspond to the true physics of the system. Previously, when presenting the PSSs in Fig. \ref{set3}(a-b) we have seen that the entire area of the phase plane defined by the ZVC is growing rapidly as the nucleus becomes more massive. Therefore, since the area of the phase plane changes significantly it gives the wrong impression regarding the evolution of chaos. This becomes clear in Fig. \ref{Mnevol2D}b where the evolution of $< \rm LCE >$ as a function of $M_n$ is presented. We see, that as the value of $M_n$ increases leading to more massive nucleus the degree of chaos also increases rapidly following an exponential trend. Here, we would like to clarify that the contradiction between Figs. \ref{Mnevol2D}a and \ref{Mnevol2D}b does nor weakens the diagnostics of chaos. On the contrary, it points out a significant property of the dynamical system, regarding the correlation between the amount and the degree of chaos. In particular, we see that as the nucleus gains more mass thus becoming more massive, the regions on phase plane may become smaller but at the same time the degree of chaos exhibits a considerable increase. In other words, the more confined are the chaotic areas in the PSS the stronger is the chaotic nature of the orbits when $M_n$ varies. Taking into account all the above-mentioned results, we may conclude that the mass of nucleus affects drastically the chaotic phenomena in our model. In particular, the more massive is the spherical nucleus the more chaos we should observe.

\section{Numerical results for the 3D system}
\label{3dsys}

In this Section, we will try to investigate the regular or chaotic nature of motion in the 3D Hamiltonian system described by Eq. (\ref{ham}). In this case, the PSS is four-dimensional and thus, not so useful as in the 2D system. Therefore, in order to keep things simple, we shall use our experience gained from the study of the two-dimensional system, in order to obtain a clear picture regarding the properties of motion in the three-dimensional model. Let us start with initial conditions on a 4D grid of the PSS. In this way, we find again regions of order and chaos, which may be visualized, if we restrict our study to a subspace of the whole 6D phase space. We consider orbits with initial conditions $(x_0,z_0,p_{x0})$, $y_0 = p_{z0} = 0$, while the initial value of $p_{y0}$ is always obtained from the energy integral (\ref{ham}). In order to maximize the accuracy of our numerical results, we use apart for the LCE, a much more efficient chaos indicator which is the SALI \citep{S01}. In particular, we define a value of $z_0$ which is kept constant and then we calculate both LCE and SALI of the 3D orbits with initial conditions $(x_0,p_{x0})$, $y_0 = p_{z0} = 0$. Thus, we are able to construct again a 2D plot depicting the $(x,p_x)$ plane but with an additional value of $z_0$. All the initial conditions of the 3D orbits lie inside the limiting curve defined by
\begin{equation}
f_3(x,p_x;z_0) = \frac{1}{2}p_x^2 + V(x;z_0) = E,
\label{ZVC3}
\end{equation}
where for convenience we take $E = h_2$.

Our extensive numerical experiments indicate, that the high complexity of the 3D dynamical system does not allow us to obtain general conclusions as we did in the previous Section. Nevertheless, by confining our study to specific levels of $z_0$ we can at least shed some light on the properties of motion of the 3D system.

\begin{figure*}
\centering
\resizebox{\hsize}{!}{\includegraphics{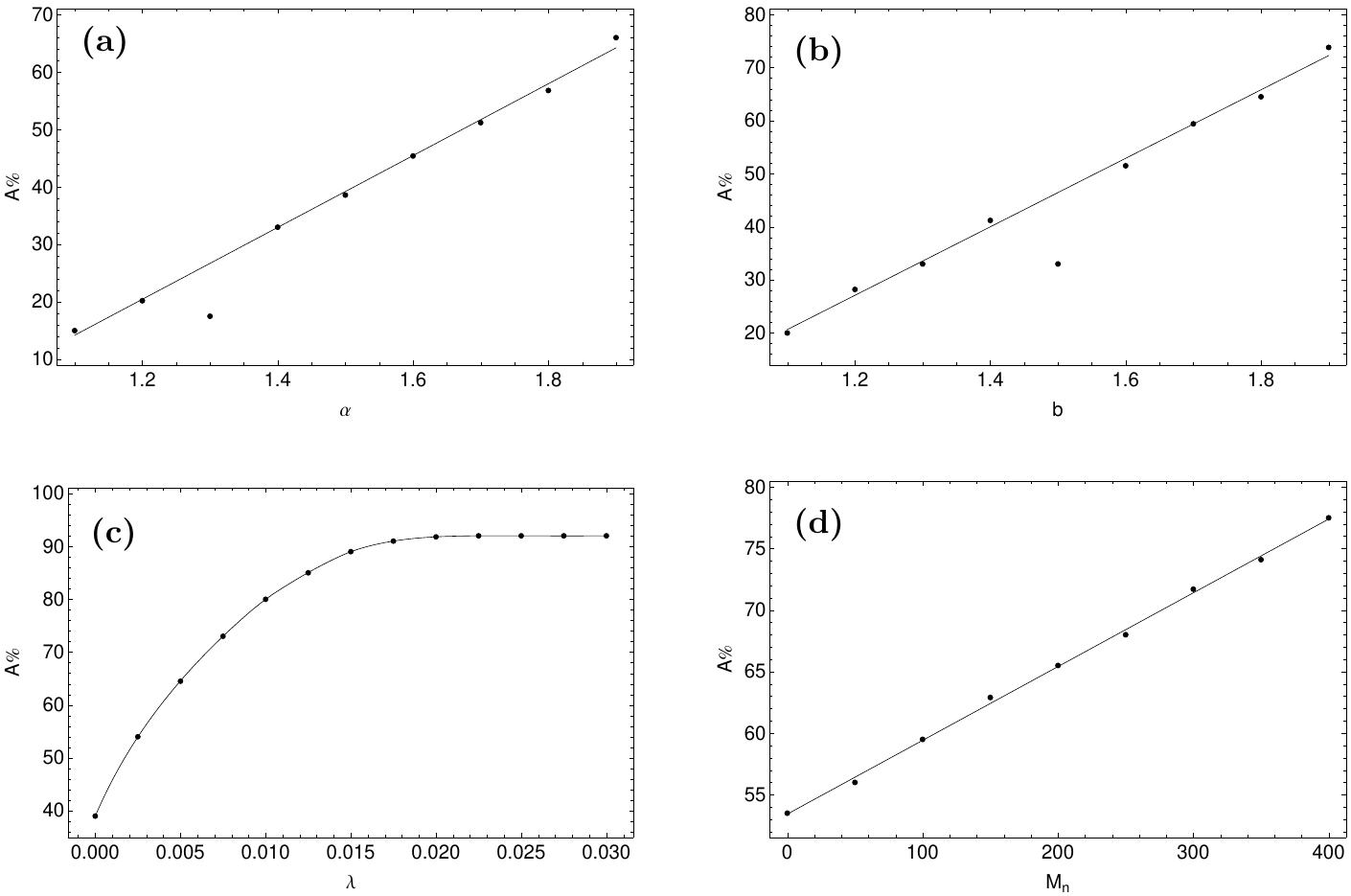}}
\caption{(a-d): Correlation between the chaotic percentage A\% of the 3D orbits and (a-upper left): flattening parameter $\alpha$, (b-upper right): flattening parameter $b$, (c-lower left): internal perturbation $\lambda$ and (d-lower right): mass of the nucleus $M_n$. More details are given in the text.}
\label{Aevol3D}
\end{figure*}

\begin{figure*}
\centering
\resizebox{\hsize}{!}{\includegraphics{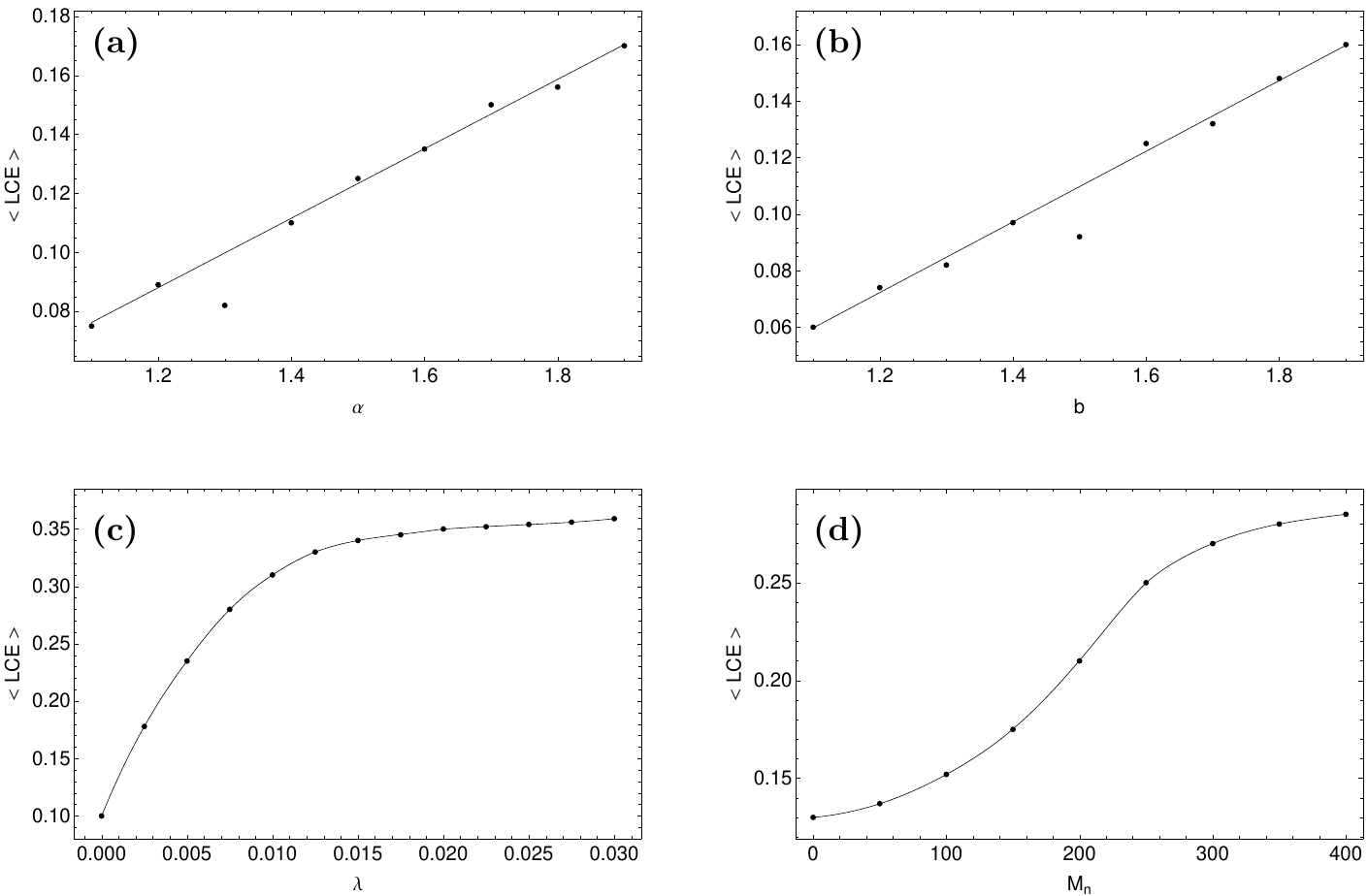}}
\caption{(a-d): Relationship between the degree of chaos of the 3D orbits, expressed by $<\rm LCE >$ and (a-upper left): flattening parameter $\alpha$, (b-upper right): flattening parameter $b$, (c-lower left): internal perturbation $\lambda$ and (d-lower right): mass of the nucleus $M_n$. More details are given in the text.}
\label{LCEevol3D}
\end{figure*}

Following the method described in the previous Section, we shall try to connect the amount and the degree of chaos with the variable parameters of the 3D dynamical system, that is the internal perturbation $\lambda$, the flattening parameters $\alpha$ and $b$ and the mass of the nucleus $M_n$. In Fig. \ref{Aevol3D}a we present the evolution of the chaotic percentage A\% of the 3D orbits as a function of the flattening parameter $\alpha$, when $\lambda = 0$, $M_n = 100$, $b = 1.3$ and $E = 500$. The initial value of $z_0$ for all the tested 3D orbits is 0.1. We observe, that the chaotic percentage increases almost linearly with increasing $\alpha$. However, when $\alpha = b = 1.3$ there is a sudden decrease caused probably by the partial symmetry $(\alpha = b)$ of the 3D system. Fig. \ref{Aevol3D}b depicts the the evolution of the chaotic percentage A\% of the 3D orbits as a function of the flattening parameter $b$, when $\lambda = 0$, $M_n = 100$, $\alpha = 1.5$ and $E = 500$. Here, the initial value of $z_0$ is 0.15. Again, the relationship between $b$ and A\% proves to be linear. Once more, when $\alpha = b = 1.5$ we observe an abrupt reduce of the chaotic percentage. The correlation between the external perturbation $\lambda$ and the chaotic percentage A\% is given in Fig. \ref{Aevol3D}. In this case, $M_n = 100$, $\alpha = 1.5$, $b = 1.3$, $E = 500$, while $z_0 = 0.2$. It is evident, that for small values of $\lambda$ the chaotic percentage increases rapidly. On the other hand, when $\lambda > 0.015$ the value of A\% remains almost constant. Finally, in Fig. \ref{Aevol3D}d we see how the mass of the nucleus $M_n$ influences the chaotic percentage of the 3D orbits. Here, $\lambda = 0.01$, $\alpha = 1.5$, $b = 1.3$, $E = 500$, while $z_0 = 0.15$. We observe, that the more massive is the nucleus the greater is the chaotic percentage.

In Fig. \ref{Aevol3D}(a-d) we presented the correlations between the amount of chaos, expressed by the chaotic percentage A\%, and the variable parameters of the dynamical system. Similarly, in Fig. \ref{LCEevol3D}(a-b) we may observe how these variable parameters influence the degree of chaos. We see that the evolution of the degree of chaos, expressed by the average value of the LCE, is quite similar to the evolution of the amount of chaos. In each case, the values of all the parameters are as in Fig. \ref{Aevol3D}. Once more, when the 3D dynamical system obtains a partial symmetry, that is when $\alpha = b$, the degree of chaos exhibits a sudden decrease. An interesting plot is shown in Fig. \ref{LCEevol3D}d, where the connection between $M_n$ and $< \rm LCE >$ is given. We observe, that when $M_n < 250$ the degree of chaos increases rapidly, while for larger values of the mass of the nucleus the increase is performed following a much smaller rate. However, we should point out, that in this case the evolution of the degree of chaos does not obey the linear law we encountered in Fig. \ref{Aevol3D}d.

\begin{figure}
\includegraphics[width=\hsize]{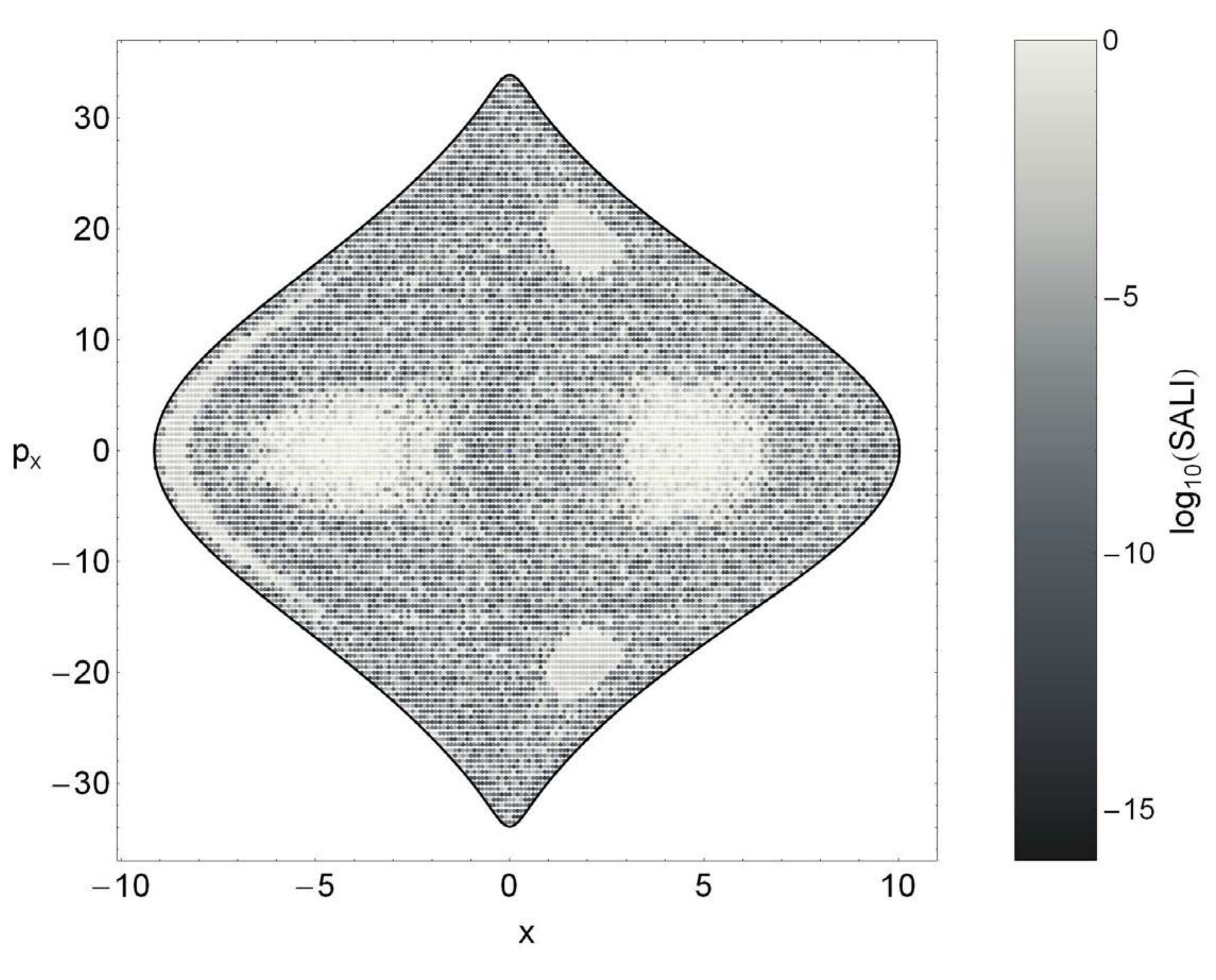}
\caption{Regions of different values of the logarithm of SALI on the $(x,p_x)$ projection of the 4D phase space when $z_0 = 1$, $\lambda = 0.01$, $M_n = 100$, $\alpha = 1.5$, $b = 1.1$ and $E = 500$. Light grey colors correspond to chaotic motion, while dark grey colors indicate ordered motion.}
\label{grid3D}
\end{figure}

\begin{figure}
\includegraphics[width=\hsize]{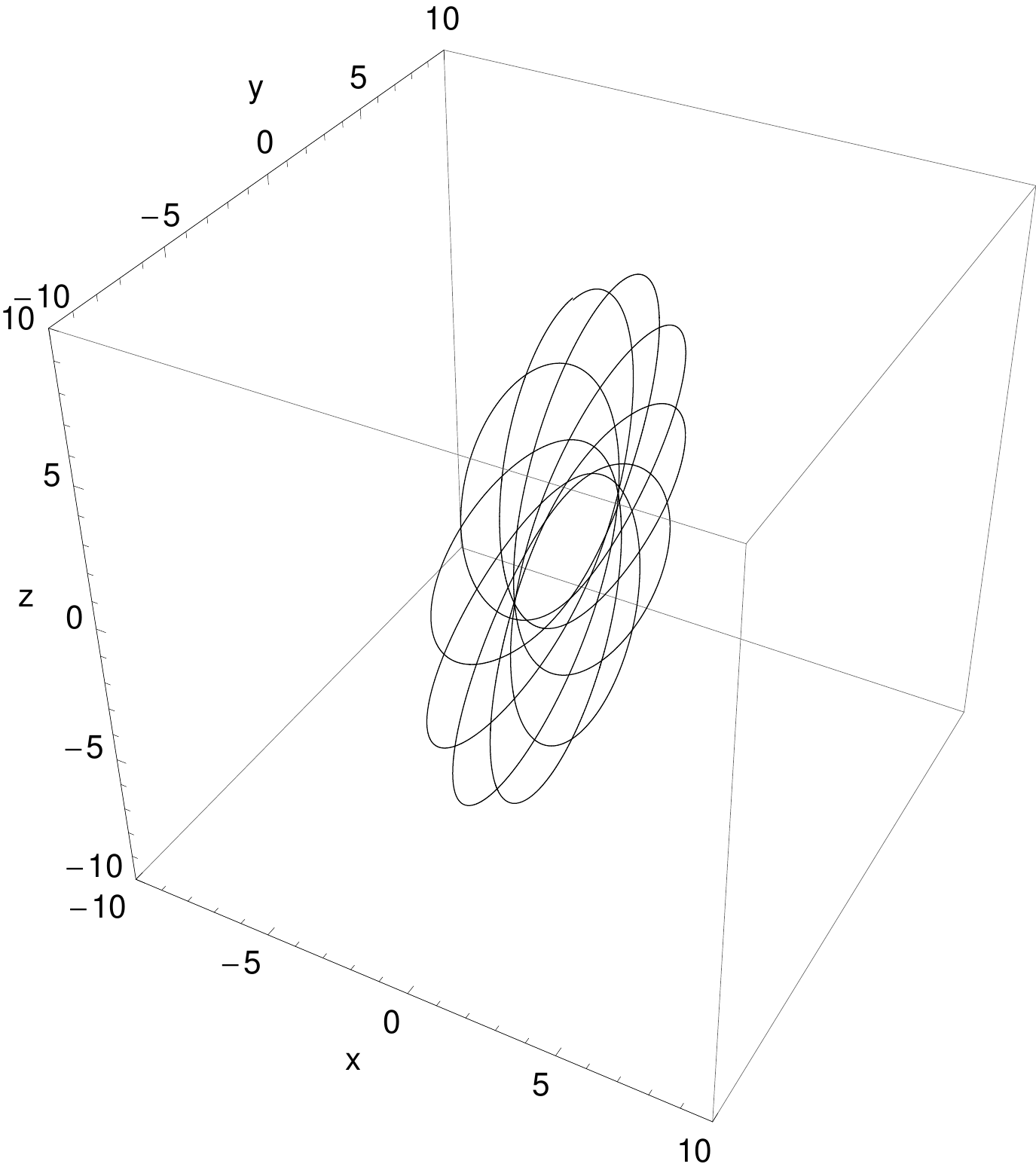}
\caption{A 3D periodic orbit circulating perpendicularly to the galactic plane. The initial conditions are $(x_0,z_0,p_{x0}) = (0.1,8.5,0)$. We point out, that the particular initial conditions $(x_0,p_{x0})$ correspond to a 2D chaotic orbit. More details are provided in the text.}
\label{orbnice}
\end{figure}

After conducting numerous numerical experiments in the 3D system, for several values of $z_0$ we arrived at the following conclusions:

(1). It was observed, that 3D orbits with initial conditions $(x_0,z_0,p_{x0})$ such as $(x_0,p_{x0})$ is a point at the chaotic regions of Figs. \ref{set1}, \ref{set2} and \ref{set3} remain chaotic only when $z_0 < 0.92$. For larger values of $z_0$ the nature of this type of orbits is inconclusive. This is true, because as we proceed to larger values of $z_0$, the islands of stability in the 4D phase space begin to destabilize and lose their well defined structure. Therefore, the initial conditions correspond to ordered or chaotic 3D orbits are completely delocalized and randomly scattered thus, preventing us from drawing safe conclusions. In Fig. \ref{grid3D} we present a grid of initial conditions corresponding to the $(x,p_x)$ projection of the 4D phase space when $z_0 = 1$. The values of the parameters are: $M_n = 100$, $\alpha = 1.5$, $b = 1.1$, $\lambda = 0.01$ and $E = 500$. The values of the logarithm of the SALI are plotted using different shades of grey. We clearly distinguish regions of regular motion indicated by dark grey colors. On the other hand, the initial conditions corresponding to chaotic motion are scattered all over the $(x,p_x)$ plane without forming a unified chaotic sea. The structure of the 4D phase space is much more complicated when $z_0 > 1$. In fact, for large values of $z_0$ the initial conditions corresponding to regular orbits are also delocalized and therefore, there is no way to have a clear picture of the 4D phase space. Such a characteristic example is given in Fig. \ref{orbnice}. Here $M_n = 400$, $\lambda = 0.01$, $\alpha = 1.5$, $b = 1.1$ and $E = 500$. The initial conditions $(x_0 = 0.1,p_{x0} = 0)$ correspond to a chaotic 2D orbit according to Fig. \ref{set3}b. However, if we use a relatively large value of $z_0$ $(z_0 = 8.5)$, we see that these initial conditions now correspond to 3D periodic orbit circulating perpendicularly to the galactic plane!

\begin{figure*}
\centering
\resizebox{\hsize}{!}{\includegraphics{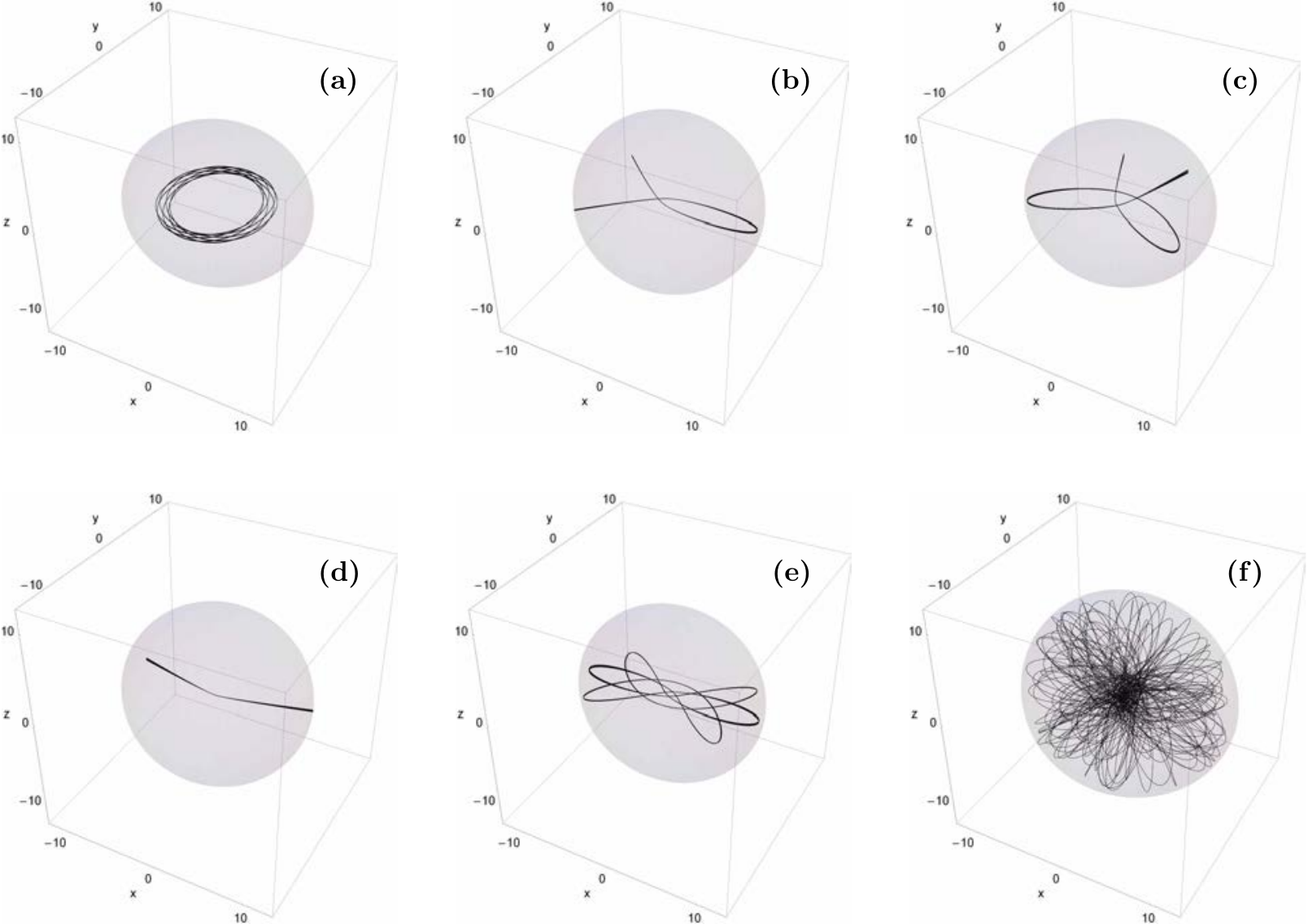}}
\caption{(a-f): Six representative orbits of the 3D dynamical system. The values of all the parameters and the initial conditions of the orbits are given in the text.}
\label{orbs3D}
\end{figure*}

(2). It was found, that the regular or chaotic nature of 3D orbits with initial conditions $(x_0,z_0,p_{x0})$ such as $(x_0,p_{x0})$ is a point at the regular regions of Figs. \ref{set1}, \ref{set2} and \ref{set3} depends strongly on the initial value of $z_0$. Orbits with low values of $z_0$ remain regular in the 3D space, while for large values of $z_0$ they alter their character and become chaotic. The general conclusion, which is based on the results derived from a large number of tested 3D orbits is that orbits with values of $z_0 < 0.82$ remain regular, while orbits with values of $z_0 \geq 0.82$ should be chaotic. The particular threshold value of $z_0$ is in fact an average value which applies to all kind of 3D orbits with initial conditions $(x_0,p_{x0})$ which correspond to regular 2D orbits. We did not feel that it was necessary to try to calculate the values of $z_{min}$ for each regular region of the 2D system corresponding to secondary resonances, which are represented by multiple sets of islands of invariant curves in the $(x,p_x)$ phase plane. The term $z_{min}$ indicates the average minimum values of $z_0$ for which the nature of a 3D orbit changes from regular to chaotic.

In Fig. \ref{orbs3D}(a-f) we present six 3D orbits of the dynamical system. We must note, that in all 3D orbits shown in Fig. \ref{orbs3D} the initial conditions $(x_0,p_{x0})$ and the values of the variable parameters are as in the corresponding 2D orbits presented in Fig. \ref{orbs2D}(a-f), while the initial value of $z_0$ is 0.15, apart from Fig. \ref{orbs3D}f where $z_0 = 1$. We observe, that all the regular 3D orbits stay relatively close to the galactic plane. The outermost gray surface which surrounds the 3D orbits is the limiting surface of the 3D space and can be obtained using the following equation
\begin{equation}
f_4(x,y,z) = V(x,y,z) = E.
\label{ZVC4}
\end{equation}

\section{Local integral of 3D motion}
\label{LocInt}

The phase space of a conservative system of three degrees of freedom has six dimensions, i.e. in Cartesian coordinates $(x,y,z,p_x,p_y,p_z)$. For a given value of the energy integral, a trajectory lies on a five-dimensional manifold. In this manifold, the surface of the section is four-dimensional. This does not allow us to visualize and interpret directly the structure and the properties of the phase space in dynamical systems of three degrees of freedom. One way to overcome this problem is to project the surface of the section to space with lower dimensions. In fact, we will apply the method introduced in \citealp{P84} (see also \citealp{RP01}). We take sections in the plane $y = 0$, $p_y > 0$ of 3D orbits, whose initial conditions differ from the plane parent periodic orbits only by the $z$ component. The set of the resulting four-dimensional points in the $(x,p_x,z,p_z)$ phase space is projected on the $(z,p_z)$ plane. If the projected points lie on a well-defined curve, we call it an ``invariant curve", then the motion is regular, while if not, the motion is chaotic. The projected points on the $(z,p_z)$ plane show nearly invariant curves around the periodic points at $z = 0$, $p_z = 0$, as long as the coupling is weak. When the coupling is stronger, the corresponding projections on the $(z,p_z)$ plane displays an increasing departure of the plane periodic point.

\begin{figure*}
\centering
\resizebox{\hsize}{!}{\includegraphics{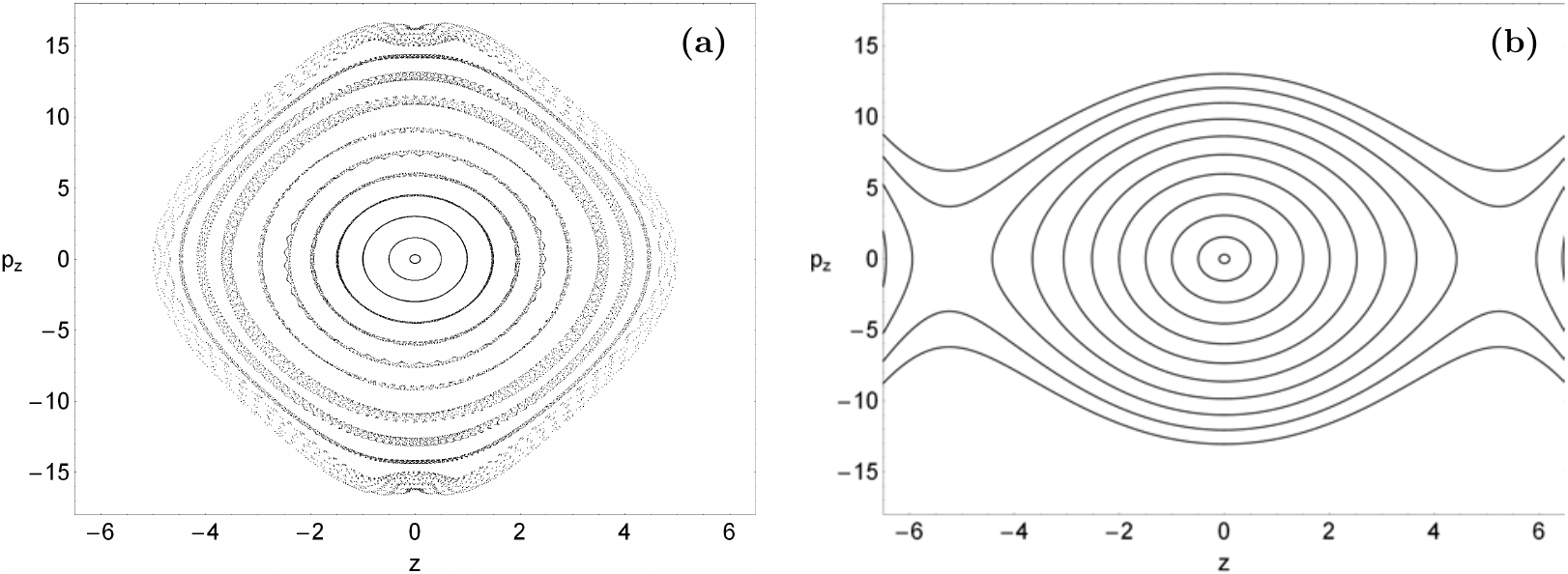}}
\caption{(a-b): (a-left): Projection of the sections of 3D orbits with the plane $y = 0$ when $p_y > 0$. The set of the four-dimensional points $(x,p_x,z,p_z)$ is projected on the $(z,p_z)$ plane. (b-right): The curves $I_z = const$ obtained theoretically when $b = 1.7$.}
\label{zpz}
\end{figure*}

Fig. \ref{zpz}a shows such``invariant curves" for orbits starting near the regular region on the right side of Fig. \ref{set3}b, when the flattening parameter has the value $b = 1.7$. In order to obtain the results shown in Fig. \ref{zpz}a we took the point $(x_0,p_{x0}) = (5.7,0)$ representing approximately the position of the periodic orbit on the $(x,p_x)$, $y_0=0, p_{y0} > 0$, phase plane and a set of values of $z_0$ = (0.1,0.5,1,1.5,2,2.5,3,3.5,4,4.5,5). Note, that for small values of $z_0$ the motion is regular, while for larger values of $z_0$ the motion is chaotic. Numerical calculations, not given here, suggest that the above method can be applied in all regular regions around the stable periodic points. However, we must emphasize that the results presented in Fig. \ref{zpz}a are rather qualitative and can be considered as an indication that the transition from regularity to chaos in 3D orbits occurs as the value of $z_0$ increases.

In what follows, we shall try to explain theoretically the numerical outcomes presented in Fig. \ref{zpz}a. We consider a point $P_0 = (x_0,y_0,z_0,p_{x0},p_{y0},p_{z0})$ in the phase space where $y_0 = 0, p_{x0} = 0$, $(x_0,0)$ being the position of the 2D periodic orbit on the $(x,p_x)$ phase plane, $z_0 = z$, $p_{z0} = p_z$, where $z$ and $p_z$ are considered as variables. The value of $p_{y0}$ is found from the energy integral (\ref{ham}). Near point $P_0$ the Hamiltonian (\ref{ham}) can be written as
\begin{eqnarray}
H &=& \frac{1}{2}\left(p_{y0}^2 + p_z^2\right) + \frac{\upsilon_0^2}{2}\ln \left(x_0^2 - \lambda x_0^3 + b z^2 + c_b^2\right) \nonumber \\
&-& \frac{M_n}{\left(x_0^2 + z^2 + c_n^2\right)^{1/2}} = E,
\label{hamP0}
\end{eqnarray}
where $z$ and $p_z$ are small compared to the values of $x_0$ and $p_{y0}$. We can rewrite (\ref{hamP0}) as
\begin{eqnarray}
H &=& \frac{1}{2}p_z^2 + \frac{\upsilon_0^2}{2}\ln\left(1 + \frac{b z^2}{A}\right) - \frac{M_n}{B^{1/2}}\left(1 + \frac{z^2}{B}\right)^{-1/2} \nonumber \\
&=& E -\frac{1}{2}p_{y0}^2 - \frac{\upsilon_0^2}{2}\ln A,
\label{hamfact}
\end{eqnarray}
where $A = x_0^2 -\lambda x_0^3 + c_b^2$ and $B = x_0^2 + c_n^2$. Next, we expand (\ref{hamfact}) in a Taylor series near the point $(z,p_z) = (0,0)$ and keeping terms up to the fourth degree in $z$ we find
\begin{eqnarray}
\frac{1}{2}p_z^2 + \frac{1}{2}\left(\frac{\upsilon_0^2 b}{A} + \frac{M_n}{B^{3/2}}\right)z^2 - \frac{1}{4}\left(\frac{\upsilon_0^2 b^2}{A^2} + \frac{3M_n}{2B^{5/2}}\right)z^4 \nonumber \\
= E - \frac{1}{2}p_{y0}^2 - \frac{\upsilon_0^2}{2}\ln A + \frac{M_n}{B^{1/2}}.
\label{hamexp}
\end{eqnarray}
Since the right hand side of Eq. (\ref{hamexp}) is constant, we may rewrite this equation in the form
\begin{equation}
I_z(z,p_z) = \frac{1}{2}p_z^2 + \frac{1}{2}\omega_0^2 z^2 + \gamma z^4 = h_3,
\label{integral}
\end{equation}
where we have set
\begin{eqnarray}
\omega_0^2 &=& \frac{\upsilon_0^2 b}{A} + \frac{M_n}{B^{3/2}}, \nonumber \\
\gamma &=& -\frac{1}{4}\left(\frac{\upsilon_0^2 b^2}{A^2} + \frac{3M_n}{2B^{5/2}}\right), \nonumber\\
h_3 &=& E - \frac{1}{2}p_{y0}^2 - \frac{\upsilon_0^2}{2}\ln A + \frac{M_n}{B^{1/2}}.
\end{eqnarray}

It is evident, that $I_z$ is indeed a local integral of the 3D motion, which is valid only in the vicinity of the two-dimensional periodic orbit $(x_0,p_{x0})$, $y_0 = 0, p_{y0} > 0$, for small values of $z_0$ and $p_{z0}$. Fig. \ref{zpz}b shows the curves $I_z = const$ for the same periodic point $(x_0,p_{x0})$ = $(5.7,0)$, $y_0 = 0, p_{y0} > 0$, when $b = 1.7$ and for the same set of values of $z_0$ as in Fig. \ref{zpz}a. We observe, that the pattern is very similar to that shown in Fig. \ref{zpz}a. Note, that for large values of $z_0$ the curves are not closed thus, implying chaotic motion. What really happens, is that for large values of $h_3$, the surface (\ref{integral}) develops saddle points. As saddle points are related to instability and chaotic motion in general we may conclude, that this can be considered as an indication for the transition from regularity to chaos. Therefore, using integral (\ref{integral}) we can obtain theoretically and with sufficient accuracy the orbital structure of the dynamical system.

\section{Linking theory with observational data}
\label{obser}

\begin{figure*}
\centering
\resizebox{\hsize}{!}{\includegraphics{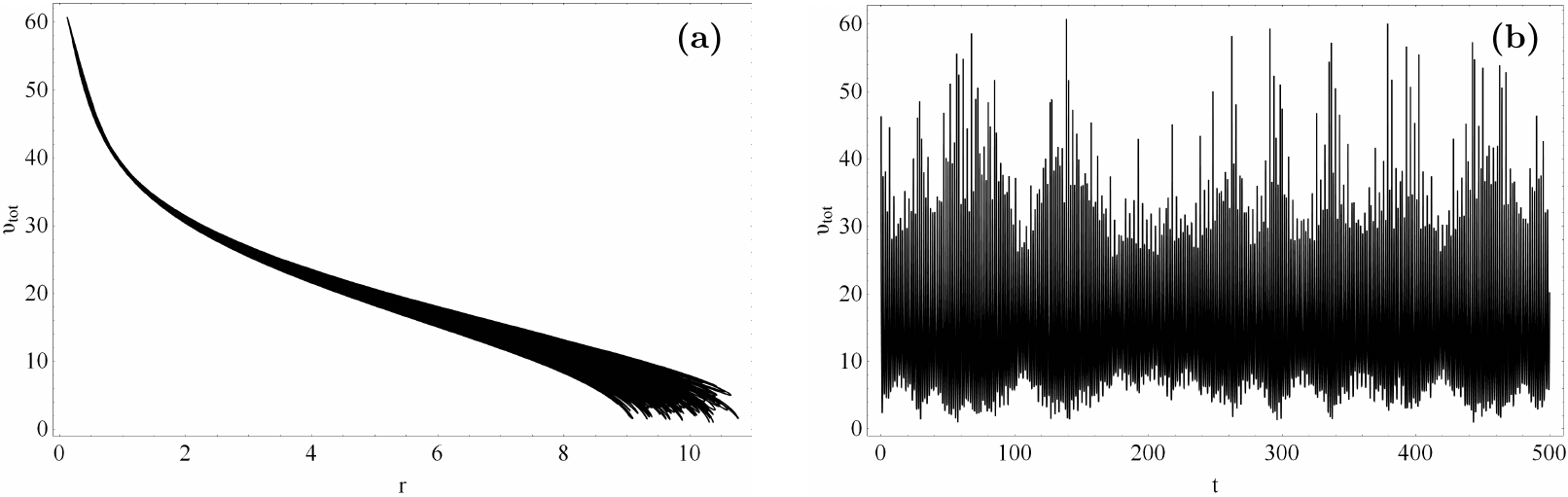}}
\caption{(a-b): A plot of the total velocity $\upsilon_{tot}$ as a function of (a-left): the distance from the galactic center $r$ and (b-right): time $t$.}
\label{vel}
\end{figure*}

In this Section, we shall try to connect some of our numerical outcomes with data derived from observations. Fig. \ref{vel}a shows the total velocity of the star as a function of the distance $r = \sqrt{x^2 + y^2 + z^2}$ from the center of the galaxy, while Fig. \ref{vel}b shows the velocity profile, that is the total velocity $\upsilon_{tot} = \sqrt{\upsilon_x^2 + \upsilon_y^2 + \upsilon_z^2}$ as a function of time for the chaotic orbit shown in Fig. \ref{orbs3D}f. There are two interesting things that should be pointed out: (i) the velocity profile shows asymmetries and abrupt changes, when approaching the nucleus and (ii) the star moves at high velocities near the central nucleus, while far from the nucleus the motion is made at low velocities. The above outcomes are in agreement with the results given by \citealp{G02}, where in regions with significant chaos one should expect high velocities and asymmetries in the velocity profile.

It is also interesting to note, that our BL Lac model is in an excellent agreement with the linear relation between the mass of the nucleus $M_n$ and the velocity dispersion $\sigma_e$ or $\nu_{rms}$ in the central parts of the galaxy \citep{FM00}. This relation reads
\begin{equation}
\log M_n = 4.8\left(\pm 0.54\right)\log \sigma_e - 2.9\left(\pm 1.3\right).
\label{vrms}
\end{equation}
Using the value $\sigma_e = 480$km/s (see Fig. \ref{set3}b) we find through Eq. (\ref{vrms}), a mean value for the mass of the galactic nucleus which equals to $< M_n >$ = 9.33 $\times$ $10^9$M$_{\odot}$. This particular value is in excellent agreement with the mass of nucleus $M_n = 400$ M.U = 9.3 $\times$ $10^9$M$_{\odot}$ used in our model.

Moreover, we can compare the maximum theoretical velocity with that obtained from observational data. That can be obtained if we set $y = p_z = 0$ in the energy integral (\ref{ham}). Then we have
\begin{equation}
\frac{1}{2}\left(p_x^2 + p_y^2\right) + V(x;z_0) = E,
\label{hamvel}
\end{equation}
where $z_0$ is the initial value corresponding to the particular 3D orbit. The maximum $p_x$ velocity $(\upsilon_{max})$ occurs on the limiting surface when $x = y = p_y = p_z =0$. Therefore
\begin{eqnarray}
\upsilon_{max} &=& \sqrt{2\left(E - V(z_0)\right)} \nonumber \\
&=& \sqrt{2\left(E - \frac{\upsilon_0^2}{2} \ln \left(b z_0^2 + c_b^2\right) + \frac{M_n}{\sqrt{z_0^2 + c_n^2}}\right)}.
\label{vmax}
\end{eqnarray}

From Eq. (\ref{vmax}) it is evident, that the maximum velocity $\upsilon_{max}$ increases as the scale length of the nucleus decreases, when all the other parameters are kept constant. In other words, higher velocities expected in galaxies with dense nuclei. For values $M_n = 400$, $b = 1.7$, $E = 500$ and for the initial value $z_0 = 1$ of the 3D chaotic orbit shown in Fig. \ref{orbs3D}f we find that $\upsilon_{max} = 383$ km/s. This value is very close to the maximum velocity observed by \citealp{B03} which was found equal to 370 km/s. Therefore, we may say that our three-dimensional dynamical model is a realistic model describing in a satisfactory way the properties of motion in a BL Lac active galaxy, since its theoretical outcomes are in sufficient agreement with related observational data.

\section{Discussion and conclusions}
\label{disc}

The Hubble Space Telescope (HST) and ground-based observations show that BL Lacs are distant galaxies with active nuclei. Today, it is clear that almost all BL Lac host galaxies are luminous ellipticals (see \citealp{USO00}). During the last years, a large amount of observational data provided a better and much more detailed picture of these active galaxies \citep[see, e.g.][]{B02,CG02,FKT02,VTN03} and also \citep{FCF04,BCC05,NTV06,ZZB07}. Therefore, all these observational data make the theoretical study of active galaxies hosting BL Lacs both interesting and challenging task.

In the present article, we constructed a three-dimensional (3D) dynamical model in order to study the motion in a galaxy hosting a BL Lac object. We believe that this model it is not only an extension in the 3D space of the two-dimensional (2D) model presented in Paper I, but also has as a target a better and more detailed description of the orbital behavior in galaxies hosting BL Lacertae objects. A galaxy hosting a BL Lac object is undoubtedly a very complex entity and, therefore, we need to assume some necessary simplifications and assumptions in order to be able to study the orbital behavior of such a complicated stellar system. Thus, our model is simple and contrived, in order to give us the ability to study different aspects of the dynamical model. Nevertheless, contrived models can provide an insight into more realistic stellar systems, which unfortunately are very difficult to be studied if we take into account all the astrophysical aspects. On the other hand, self-consistent models are mainly used when conducting N-body simulations. However, this is entirely out of the scope of the present paper. Once again, note that the simplicity of our model is necessary; otherwise it would be extremely difficult, or even impossible, to apply the extensive and detailed dynamical study presented in this study. Similar gravitational models with the same limitations and assumptions were used successfully several times in the past in order to investigate the orbital structure in much more complicated galactic systems \citep{Z12b,Z13}.

In order to make things simple, we started our investigation from the 2D model using the numerical outcomes as a starting point. Next we focused our study to the three-dimensional (3D) system, where the corresponding PSS is four-dimensional and thus, cannot be visualized directly. In an attempt to overcome this drawback, we used our experience gained from the study of the 2D system, to obtain a clear picture regarding the properties of motion in the three-dimensional model. In order to optimize the accuracy of our numerical results, we combined two well-tested and efficient chaos indicators (LCE and SALI) so as to distinguish between regular and chaotic motion. In particular, we constructed 2D grids of initial conditions $(x_0,p_{x0})$ in which we computed the LCE and SALI of the 3D orbits for several predefined values of $z_0$. Remember, that for all orbits we took $y_0 = p_{z0}=0$, while the value of $p_{y0}$ was found from the energy integral (\ref{ham}). Our extensive numerical experiments revealed, that the high complexity of the 3D system prevent us from obtaining general conclusions as in the case of the 2D model. However, by confining our study to specific levels of $z_0$ we managed to shed some light on the properties of motion in the 3D system.

Several correlations between the basic dynamical parameters of the galaxy and both the degree and amount of chaos were found to exist. The main outcomes of our research can be summarized as follows:
\begin{enumerate}
  \item It was observed, that the presence of a massive and dense nucleus at the center of the galaxy increases the relative percentage of the chaotic orbits in the phase plane. This conclusion fully agrees with the findings presented in \citep{Z12a} and \citep{ZC13}, where we investigated the influence of the spherical nucleus in an axially symmetric galactic gravitational model with an additional disk-halo component.
  \item A significant increase regarding the allowed velocities of stars near the central region of the galaxy was measured in the case where a dense and massive nucleus is present. Moreover, asymmetries and abrupt changes were detected in the velocity profile of chaotic orbits when approaching the spherical nucleus.
  \item A linear relationship between both the percentage and the degree of chaotic orbits and the flattening parameters $\alpha$ and $b$ was found in both the 2D and 3D systems. In particular, the more flattened is the galaxy along the $y$ and $z$ axes the more chaos should we observe.
  \item The parameter $\lambda$ which determines the strength of the internal perturbation, or in other words the deviation from axial symmetry affects greatly the amount of chaotic orbits in the galaxy. Specifically, the percentage of chaotic orbits grows rapidly with increasing perturbation.
  \item One of the most influential factors which determines the regular or chaotic character of 3D orbits is the initial value of the $z$ coordinate. Our numerical analysis suggests that in general terms orbits with low values of $z_0$ retain their 2D character (regular or chaotic). For $z_0 \gtrsim 0.8$ on the other hand, the structure of the phase space becomes very complex thus preventing us from drawing safe orbit classification.
\end{enumerate}

It was the complexity of the 3D system that forced us to develop new theoretical arguments in order to interpret and support the numerically obtained outcomes regarding 3D motion (ordered or chaotic). We found, that near the vicinity of stable periodic points, the nature of the 3D orbits can indeed be explained using a local integral of motion. This local integral of motion is in fact, the energy of a test particle (star) at the $z$ direction. A qualitative distinction between regular and chaotic orbits can be obtained by looking the projection of the 4D space at the $(z,p_z)$ plane. If the curve corresponding to a 3D orbit is closed then the motion is ordered, while if the curve is open, we have strong numerical evidence that this implies chaotic motion.

Taking into account that the present numerical outcomes are in sufficient agreement with several related data derived from observations, we may conclude that our three-dimensional dynamical model (\ref{vtot}) is indeed a realistic candidate for modeling the dynamical profile in a BL Lac active galaxy. We consider the outcomes of the present research as an initial effort in the task of exploring the orbital structure of galaxies hosting a BL Lac objects. Since our results are encouraging, it is in our future plans to study the influence of all the available parameters in a time-dependent and also rotating dynamical system. Active galaxies (AGNs) is a modern and fast developing branch of Observation Astronomy. Therefore, we hope to be able in the near future to construct much better dynamical models in order to reveal the true nature of these impressive stellar objects.

\section*{Acknowledgments}

The author would like to thank the anonymous referee for the careful reading of the manuscript and for all the aptly suggestions and comments which allowed us to improve both the quality and the clarity of our work.

\end{document}